\documentclass{PoS}

\usepackage{wrapfig}

\newcommand{\be}{\begin{equation}}
\newcommand{\ee}{\end{equation}}
\newcommand{\bea}{\begin{eqnarray}}
\newcommand{\eea}{\end{eqnarray}}

\newcommand{\Tb}{\bar{T}_b}

\title{HI Intensity Mapping with MeerKAT} 

\ShortTitle{HI Intensity Mapping with MeerKAT} 

\author{\speaker{Alkistis Pourtsidou}%
        \\ 
        School of Physics and Astronomy, Queen Mary University of London, Mile End Road, 
        London E1 4NS, United Kingdom
          \\
       Institute of Cosmology and Gravitation, University of Portsmouth,
Dennis Sciama Building, Burnaby Road, Portsmouth, PO1 3FX, United Kingdom\\ 
       E-mail: \email{a.poursidou@qmul.ac.uk}} 

\abstract{
Radio detections of the redshifted 21cm line with neutral hydrogen (HI) intensity mapping surveys can provide a new way to probe the evolving history of the Universe and tackle fundamental cosmological questions. Intensity mapping (IM) measures the fluctuations of the HI signal tracing the underlying matter distribution, and therefore can be used to reconstruct the matter power spectrum without the need to detect individual galaxies. This means that intensity mapping surveys can observe large volumes very fast, and give us access to large cosmological scales while providing excellent redshift information.
We explore the possibility of performing such a survey with the South African MeerKAT radio telescope, which is a precursor to the Square Kilometre Array (SKA). We also propose to use cross-correlations between the MeerKAT intensity mapping survey and optical galaxy surveys, in order to mitigate important systematic effects and produce robust cosmological measurements. Our forecasts show that precise measurements of the HI signal can be made in the near future. These can be used to constrain HI and cosmological parameters across a wide range of redshift. 
}

\FullConference{MeerKAT Science: On the Pathway to the SKA,\\
	25-27 May, 2016,\\
		Stellenbosch, South Africa}

\begin{document}

\section{Introduction}

Intensity mapping \cite{Chang:2007xk,Loeb:2008hg,Mao:2008ug,Peterson:2009ka,Seo:2009fq,Ansari:2011bv,Battye:2012tg,Switzer:2013ewa} is an innovative technique that uses neutral hydrogen to map the large-scale structure of the Universe in three dimensions. Intensity mapping surveys use HI as a tracer of the underlying dark matter distribution but, unlike traditional galaxy surveys, they do not detect individual galaxies; instead, they treat the 21cm sky as a diffuse background measuring the integrated intensity of the redshifted 21cm line across the sky and along redshift.

Recent studies have shown that we can deliver precision cosmology (e.g. Baryonic Acoustic Oscillations measurements) using SKA Phase 1 and the IM method \cite{Bull:2014rha, Santos:2015bsa, Pourtsidou:2015mia}. Here we will demonstrate that we can use the MeerKAT array to exploit the IM technique and constrain HI and cosmological parameters across a wide range of redshift. This will be an important step towards SKA1\_MID large sky surveys, and it will help us maximise their scientific output. 
We will first investigate what can be achieved using the auto-correlation HI clustering measurements, and then move on to cross-correlations with optical galaxy surveys. Cross-correlations between different surveys are expected to yield precise and robust cosmological constraints. They can alleviate various survey-specific challenges (foregrounds, systematics) that are expected to drop out in cross correlation.

\section{HI intensity mapping with MeerKAT}

MeerKAT\footnote{http://www.ska.ac.za/science-engineering/meerkat/} is a 64-dish SKA pathfinder on the planned site of SKA1\_MID, with 20 dishes already in place. A large sky intensity mapping survey with MeerKAT has been proposed \cite{Santos:2017qgq}. It will scan a few thousand degrees on the sky --we take $A_{\rm sky} = 4000 \; {\rm deg}^2$ here-- in approximately 5 months total observation time. The array will operate in single dish mode in order to access large (cosmological) scales, and there are two frequency bands available\footnote{Note that the survey will have to operate in one band only.}: the L band $0<z<0.58$ ($900<f<1420$ MHz), and the UHF band $0.4<z<1.45$ ($580<f<1000$ MHz). A detailed description of the noise properties of such a survey can be found in \cite{Pourtsidou:2015mia}. Note that the 
survey will also take interferometric data simultaneously, and various science cases can be achieved: cosmology, galaxy evolution, clusters, galaxy HI emission, and polarization.

\begin{figure}
\includegraphics[scale=0.5]{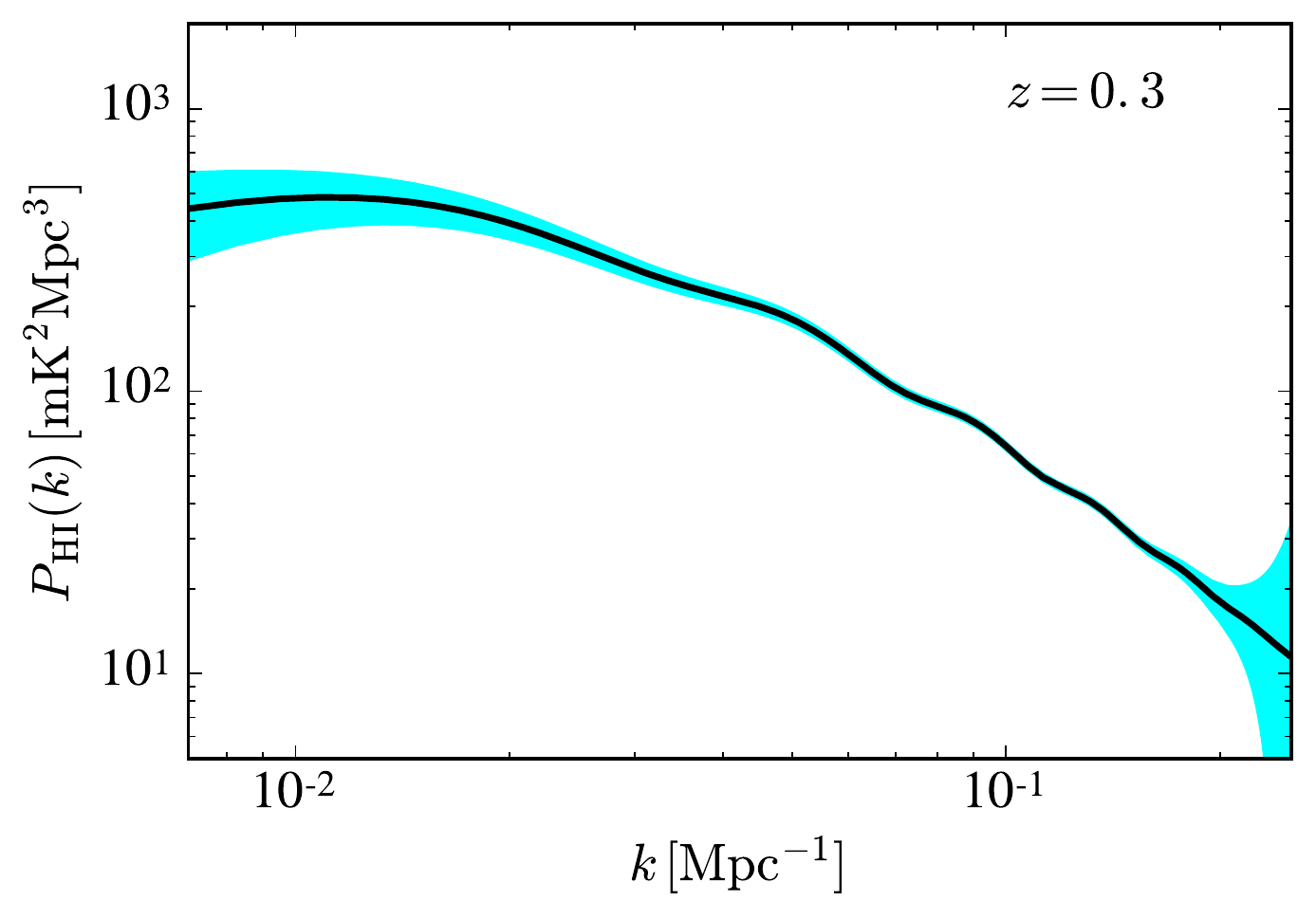}
\includegraphics[scale=0.5]{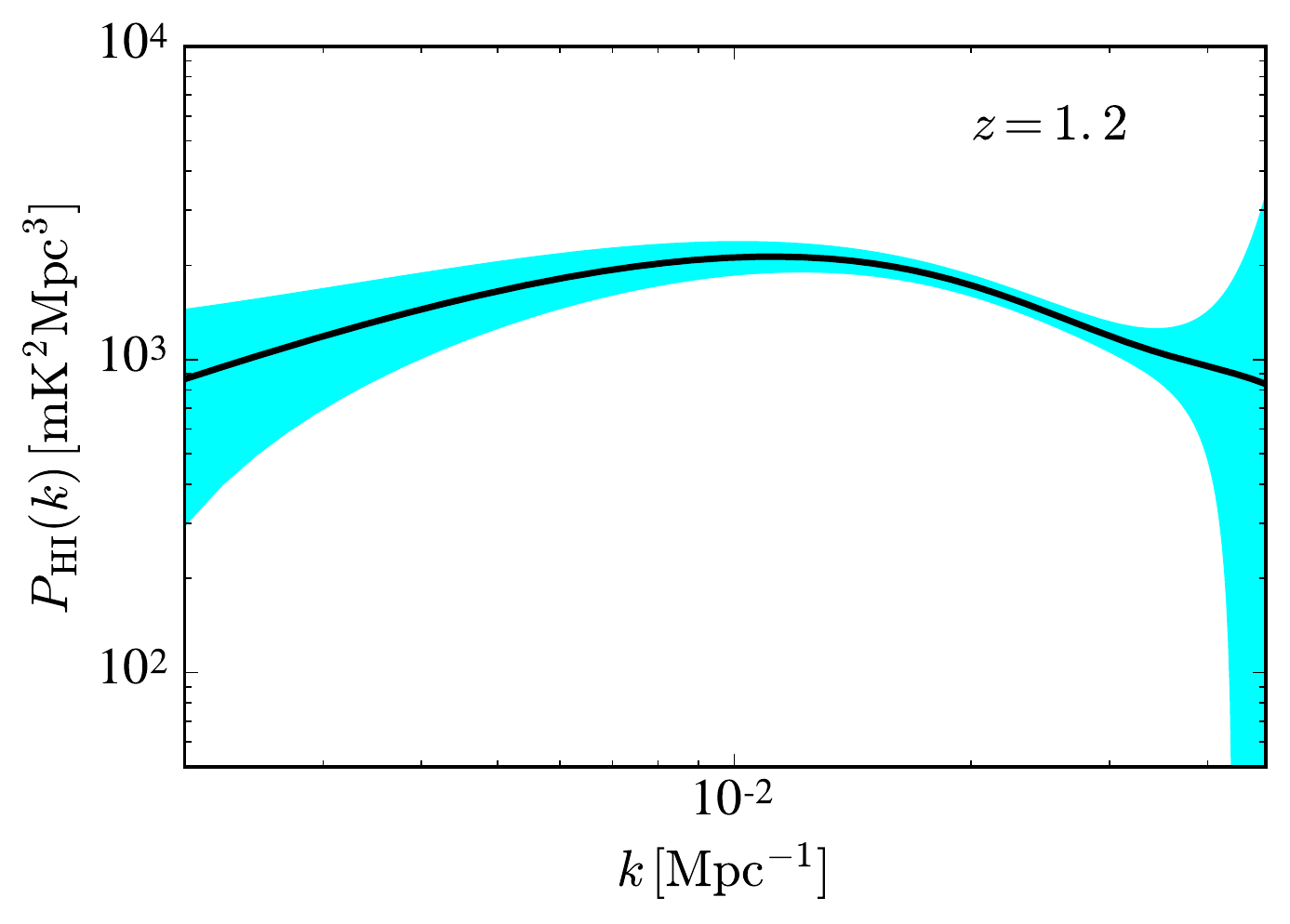}
\caption{HI detection in autocorrelation with a $4000 \; {\rm deg}^2$ survey using MeerKAT. The black solid line is the predicted power spectrum $P_{\rm HI}(k,z)$ at $z=0.3$ (left) and at $z=1.2$ (right). Note that $z=0.3$ is in the L-band while $z=1.2$ is in the UHF band. The cyan area represents the measurement errors for a total observation time of $\sim 5$ months. The width of the bins is $\Delta z = 0.1$ and we have used a $k$-binning $\Delta k = 0.01 \; {\rm Mpc}^{-1}$.}
\label{fig:PkHI}
\end{figure}

The HI power spectrum can be written as 
\be
P^{\rm HI}(k,z)=\Tb^2 b^2_{\rm HI}P(k,z) \, ,
\ee where $P$ is the matter power spectrum, and $b_{\rm HI}$ the HI bias. The mean 21cm emission brightness temperature is given by
$
\Tb(z) = 180 \Omega_{\rm HI}(z)h\frac{(1+z)^2}{H(z)/H_0} \, {\rm mK},
$ where $\Omega_{\rm HI}$ is the HI density, $H(z)$ the Hubble parameter as a function of redshift $z$, and $H_0\equiv 100h$ its value today \cite{Battye:2012tg}.
For modelling the HI density we use $\Omega_{\rm HI} \simeq 0.00048+0.00039z$, and for the HI bias we use $b_{\rm HI} \simeq 0.67+0.18z$ \cite{Bull:2014rha}. The noise power spectrum $P^{\rm N}$ is calculated using the formalism described in \cite{Pourtsidou:2015mia}. In Figure~\ref{fig:PkHI} we plot the HI power spectrum and measurements errors for two bins centred at $z=0.3$ and $z=1.2$. 

As we can see from the above plots, a very good signal-to-noise ratio can be achieved over a wide range of scales and redshifts. These measurements can be used to constrain HI and cosmological parameters. As a first approach we assume a flat $\Lambda$CDM expansion history and keep all cosmological parameters fixed to the Planck 2015 cosmology \cite{Ade:2015xua}. Then the only unknown in the HI power spectrum is the prefactor $\Omega_{\rm HI}b_{\rm HI}$. We can therefore employ the Fisher matrix formalism \cite{Tegmark:1997rp} to forecast constraints on the HI abundance and bias along redshift. Using the aforementioned MeerKAT IM survey parameters, we find $\delta(\Omega_{\rm HI}b_{\rm HI})/(\Omega_{\rm HI}b_{\rm HI})=0.005$ at $z=0.3$ (L band) and $\delta(\Omega_{\rm HI}b_{\rm HI})/(\Omega_{\rm HI}b_{\rm HI})=0.03$ at $z=1.2$ (UHF band) \cite{Pourtsidou:2016dzn}. These are more than one order of magnitude better than the currently available constraints from galaxy surveys, intensity mapping, and damped Lyman-$\alpha$ observations (see Table 2 in \cite{Padmanabhan:2014zma}). In \cite{Pourtsidou:2016dzn} a comprehensive analysis is performed and constraints are derived for multiple redshifts across the L and UHF bands. 

Including redshift space distortions (RSD) the HI power spectrum can be written as
\be
P^{\rm HI}(k,z;\mu)=\Tb^2 b^2_{\rm HI} [1+\beta_{\rm HI}(z)\mu^2]^2P(k,z) \, ,
\ee where $\mu = \hat{k} \cdot \hat{z}$ and $\beta_{\rm HI}$ is the redshift space distortion parameter equal to $f/b_{\rm HI}$ in linear theory, with $f \equiv d{\rm ln}D/d{\rm ln}a$ the linear growth rate. RSD measurements can be used to break the degeneracy between $\Omega_{\rm HI}$ and $b_{\rm HI}$ \cite{Masui:2010mp, Pourtsidou:2016dzn}. Assuming a $\Lambda$CDM expansion history, we find $\delta \Omega_{\rm HI}/\Omega_{\rm HI}=0.03$ at $z=0.3$ (L band) and $\delta \Omega_{\rm HI}/\Omega_{\rm HI}=0.09$ at $z=1.2$ (UHF band). Another approach is to assume that the mean temperature $\bar{T}_b$ has been measured using the smooth part of the HI signal \cite{Bull:2014rha}, and constrain the HI bias and various cosmological parameters. Forecasts for constraints on $\{f\sigma_8, b_{\rm HI}\sigma_8, D_{\rm A}, H\}$ using an IM survey with MeerKAT can be found in \cite{Pourtsidou:2016dzn}. For example, the (marginalised) fractional error on $f\sigma_8(z)$ is $3\%$ at $z=0.3$ (L band) and $13\%$ at $z=1.2$ (UHF band) for the survey parameters we have considered here.

\section{Cross-correlations with optical galaxy surveys}

\begin{wrapfigure}{R}{0.55\textwidth}
\centering
\includegraphics[width=0.5\textwidth]{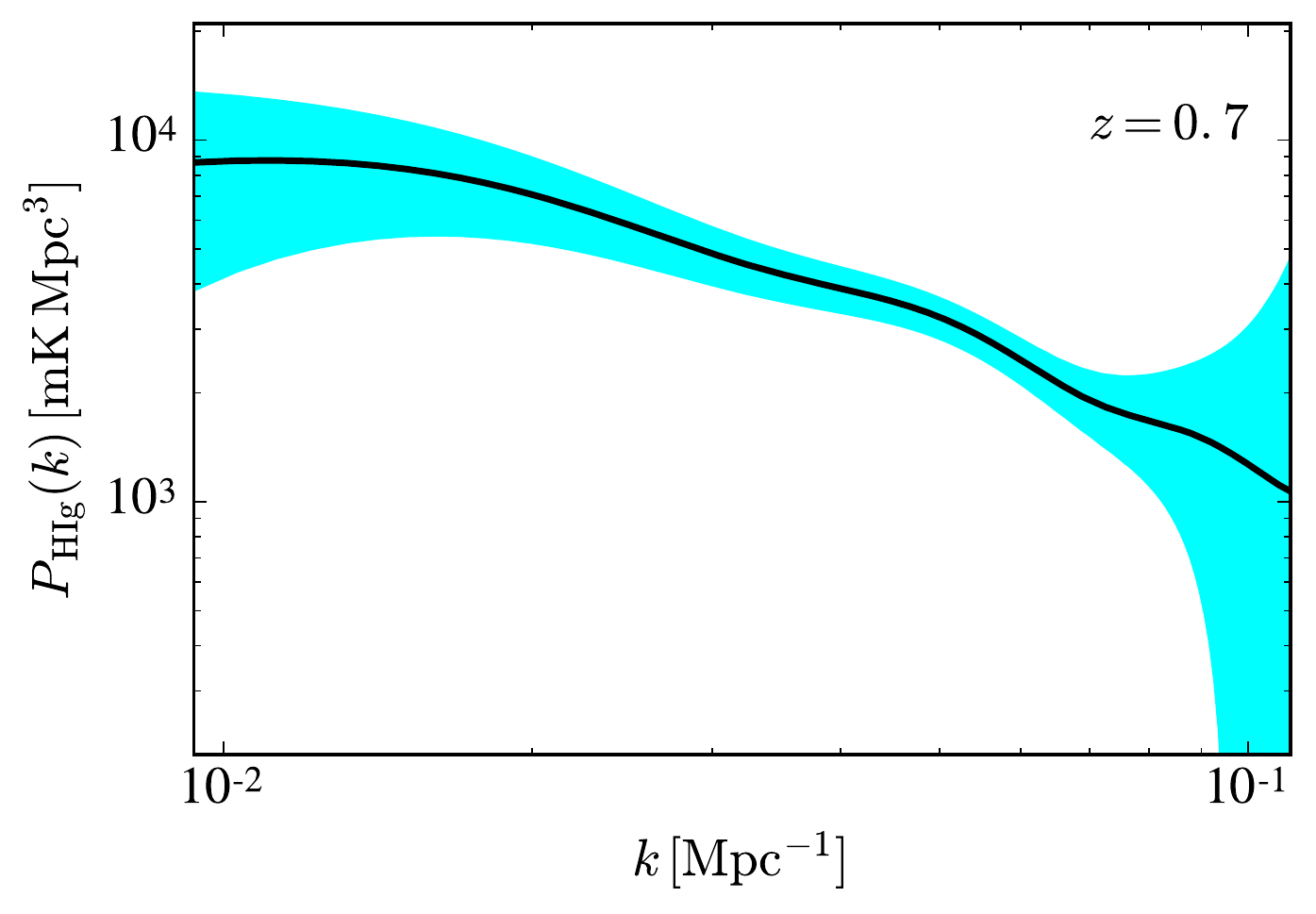}
\caption{HI detection in cross-correlation assuming the MeerKAT IM survey (UHF band) and a Euclid-like spectroscopic survey with $500 \; {\rm deg}^2$ overlap at $z=0.7$. The cyan area represents the measurement errors for a total observation time of $500$ hrs ($\Delta z = 0.1$, $\Delta k = 0.01 \; {\rm Mpc}^{-1}$).}
\label{fig:Pkcross}
\end{wrapfigure}
Cross correlations between HI intensity mapping and optical galaxy surveys can provide precise
 and robust cosmological measurements, as they have the advantage of mitigating major issues like systematics and foreground contaminants that are relevant for one type of survey but not for the other \cite{Masui:2012zc,Wolz:2015lwa}. 
In \cite{Masui:2012zc} the 21cm maps acquired at the Green Bank Telescope were combined with the WiggleZ galaxy survey \cite{Parkinson:2012vd} to constrain the quantity $\Omega_{\rm HI}b_{\rm HI}r$ --with $r$ a correlation coefficient accounting for the possible stochasticity in the galaxy and HI tracers-- at $z\sim 0.8$ with a statistical fractional error $\sim 16\%$.  The cross correlation power spectrum can be written as \cite{Masui:2012zc}
\be
P^{\rm HI,g}(k,z) = \bar{T}_b b_{\rm HI}b_{\rm g}r P(k,z) \, ,
\ee with $b_g$ the galaxy bias. Here we will consider a forthcoming spectroscopic survey with Euclid-like specifications \cite{Amendola:2016saw}, which is assumed to have a (conservative) $500 \; {\rm deg}^2$ sky overlap with MeerKAT, with the observing time reduced accordingly with respect to the full $4,000 \; {\rm deg}^2$ MeerKAT survey. The redshift overlap is $0.7<z<1.4$ (UHF band). In Figure~\ref{fig:Pkcross} we plot the cross-correlation power spectrum and measurement errors for a bin centred at $z=0.7$. 
Assuming a $\Lambda$CDM expansion history and $b_g$ measured from the galaxy survey, we find $\delta(\Omega_{\rm HI}b_{\rm HI}r)/(\Omega_{\rm HI}b_{\rm HI}r)=0.06$ at $z=0.7$ and $\delta(\Omega_{\rm HI}b_{\rm HI}r)/(\Omega_{\rm HI}b_{\rm HI}r)=0.14$ at $z=1.2$. 
Modifying our approach as before by including RSDs and considering $\bar{T}_b$ and $b_{\rm g}$ known we can constrain the growth of structure and expansion history of the Universe. In \cite{Pourtsidou:2016dzn} a comprehensive analysis is performed and constraints are derived for multiple redshifts across the L and UHF bands. Note that assuming an IM survey with SKA1-MID and a $7000 \, {\rm deg}^2$ overlap with Euclid we find a (marginalised) $4\%$ error on $f\sigma_8$ at $z=0.7$ (see \cite{Pourtsidou:2016dzn} for details).
Another possibility is to combine MeerKAT and SKA1-MID IM surveys with photometric galaxy surveys like DES\footnote{https://www.darkenergysurvey.org/} and LSST\footnote{https://www.lsst.org/}. By doing so we can detect HI clustering and weak gravitational lensing of the 21cm emission in cross-correlation \cite{Pourtsidou:2015mia}. 

\section{Conclusions}

We have shown that using the MeerKAT pathfinder we can utilise the intensity mapping method to constrain HI and cosmological parameters across a wide range of redshift. This way we can explore and maximise the scientific output of future large sky surveys with Phase 1 of the SKA. Furthermore, we have proposed to exploit cross-correlations of the 21cm intensity maps with galaxies, in order to have the advantage of reducing systematic effects. We believe that the  MeerKAT array can be at the forefront of 21cm cosmology with the next generation of radio telescopes, and we consider the auto- and cross-correlation approaches highly complementary and synergistic.

\vspace{0.5cm}

\noindent{\it Acknowledgments:} --- This work was supported by a Dennis Sciama Fellowship at the University of Portsmouth. I acknowledge the University of the Western Cape for hospitality. I would like to thank David Bacon, Robert Crittenden, Roy Maartens, Ben Metcalf, and Mario Santos for fruitful collaborations and useful discussions. 

\bibliographystyle{JHEP}
\bibliography{references_MK}

\providecommand{\href}[2]{#2}\begingroup\raggedright\begin{thebibliography}{10}

\bibitem{Chang:2007xk}
T.-C. Chang, U.-L. Pen, J.~B. Peterson and P.~McDonald, \emph{{Baryon Acoustic
  Oscillation Intensity Mapping as a Test of Dark Energy}},
  \href{https://doi.org/10.1103/PhysRevLett.100.091303}{\emph{Phys.Rev.Lett.}
  {\bfseries 100} (2008) 091303},
  [\href{https://arxiv.org/abs/0709.3672}{{\ttfamily 0709.3672}}].

\bibitem{Loeb:2008hg}
A.~Loeb and S.~Wyithe, \emph{{Precise Measurement of the Cosmological Power
  Spectrum With a Dedicated 21cm Survey After Reionization}},
  \href{https://doi.org/10.1103/PhysRevLett.100.161301}{\emph{Phys.Rev.Lett.}
  {\bfseries 100} (2008) 161301},
  [\href{https://arxiv.org/abs/0801.1677}{{\ttfamily 0801.1677}}].

\bibitem{Mao:2008ug}
Y.~Mao, M.~Tegmark, M.~McQuinn, M.~Zaldarriaga and O.~Zahn, \emph{{How
  accurately can 21 cm tomography constrain cosmology?}},
  \href{https://doi.org/10.1103/PhysRevD.78.023529}{\emph{Phys.Rev.} {\bfseries
  D78} (2008) 023529}, [\href{https://arxiv.org/abs/0802.1710}{{\ttfamily
  0802.1710}}].

\bibitem{Peterson:2009ka}
J.~B. Peterson, R.~Aleksan, R.~Ansari, K.~Bandura, D.~Bond et~al., \emph{{21 cm
  Intensity Mapping}},  \href{https://arxiv.org/abs/0902.3091}{{\ttfamily
  0902.3091}}.

\bibitem{Seo:2009fq}
H.-J. Seo, S.~Dodelson, J.~Marriner, D.~Mcginnis, A.~Stebbins et~al., \emph{{A
  ground-based 21cm Baryon acoustic oscillation survey}},
  \href{https://doi.org/10.1088/0004-637X/721/1/164}{\emph{Astrophys.J.}
  {\bfseries 721} (2010) 164--173},
  [\href{https://arxiv.org/abs/0910.5007}{{\ttfamily 0910.5007}}].

\bibitem{Ansari:2011bv}
R.~Ansari, J.~Campagne, P.~Colom, J.~L. Goff, C.~Magneville et~al., \emph{{21
  cm observation of LSS at z\textasciitilde 1 Instrument sensitivity and
  foreground subtraction}},
  \href{https://doi.org/10.1051/0004-6361/201117837}{\emph{Astron.Astrophys.}
  {\bfseries 540} (2012) A129},
  [\href{https://arxiv.org/abs/1108.1474}{{\ttfamily 1108.1474}}].

\bibitem{Battye:2012tg}
R.~Battye, I.~Browne, C.~Dickinson, G.~Heron, B.~Maffei et~al., \emph{{HI
  intensity mapping : a single dish approach}},
  \href{https://doi.org/10.1093/mnras/stt1082}{\emph{Mon. Not. Roy. Astron.
  Soc} {\bfseries 434} (2013) 1239--1256},
  [\href{https://arxiv.org/abs/1209.0343}{{\ttfamily 1209.0343}}].

\bibitem{Switzer:2013ewa}
E.~Switzer, K.~Masui, K.~Bandura, L.~M. Calin, T.~C. Chang et~al.,
  \emph{{Determination of z\textasciitilde0.8 neutral hydrogen fluctuations
  using the 21 cm intensity mapping auto-correlation}},
  \href{https://doi.org/10.1093/mnrasl/slt074}{\emph{Mon.Not.Roy.Astron.Soc.}
  {\bfseries 434} (2013) L46--L50},
  [\href{https://arxiv.org/abs/1304.3712}{{\ttfamily 1304.3712}}].

\bibitem{Bull:2014rha}
P.~Bull, P.~G. Ferreira, P.~Patel and M.~G. Santos, \emph{{Late-time cosmology
  with 21cm intensity mapping experiments}},
  \href{https://doi.org/10.1088/0004-637X/803/1/21}{\emph{Astrophys.J.}
  {\bfseries 803} (2015) 21},
  [\href{https://arxiv.org/abs/1405.1452}{{\ttfamily 1405.1452}}].

\bibitem{Santos:2015bsa}
M.~Santos et~al., \emph{{Cosmology from a SKA HI intensity mapping survey}},
  {\emph{PoS} {\bfseries AASKA14} (2015) 019}.

\bibitem{Pourtsidou:2015mia}
A.~Pourtsidou, D.~Bacon, R.~Crittenden and R.~B. Metcalf, \emph{{Prospects for
  clustering and lensing measurements with forthcoming intensity mapping and
  optical surveys}}, \href{https://doi.org/10.1093/mnras/stw658}{\emph{Mon.
  Not. Roy. Astron. Soc.} {\bfseries 459} (2016) 863--870},
  [\href{https://arxiv.org/abs/1509.03286}{{\ttfamily 1509.03286}}].

\bibitem{Santos:2017qgq}
M.~G. Santos et~al., \emph{{MeerKLASS: MeerKAT Large Area Synoptic Survey}},
  2017, \href{https://arxiv.org/abs/1709.06099}{{\ttfamily 1709.06099}},
  \href{http://inspirehep.net/record/1624378/files/arXiv:1709.06099.pdf}{http://inspirehep.net/record/1624378/files/arXiv:1709.06099.pdf}.

\bibitem{Ade:2015xua}
{\scshape Planck} collaboration, P.~A.~R. Ade et~al., \emph{{Planck 2015
  results. XIII. Cosmological parameters}},
  \href{https://doi.org/10.1051/0004-6361/201525830}{\emph{Astron. Astrophys.}
  {\bfseries 594} (2016) A13},
  [\href{https://arxiv.org/abs/1502.01589}{{\ttfamily 1502.01589}}].

\bibitem{Tegmark:1997rp}
M.~Tegmark, \emph{{Measuring cosmological parameters with galaxy surveys}},
  \href{https://doi.org/10.1103/PhysRevLett.79.3806}{\emph{Phys. Rev. Lett.}
  {\bfseries 79} (1997) 3806--3809},
  [\href{https://arxiv.org/abs/astro-ph/9706198}{{\ttfamily
  astro-ph/9706198}}].

\bibitem{Pourtsidou:2016dzn}
A.~Pourtsidou, D.~Bacon and R.~Crittenden, \emph{{HI and cosmological
  constraints from intensity mapping, optical, and CMB surveys}},
  \href{https://doi.org/10.1093/mnras/stx1479}{\emph{Mon. Not. Roy. Astron.
  Soc.} {\bfseries 470} (2017) 4251--4260},
  [\href{https://arxiv.org/abs/1610.04189}{{\ttfamily 1610.04189}}].

\bibitem{Padmanabhan:2014zma}
H.~Padmanabhan, T.~R. Choudhury and A.~Refregier, \emph{{Theoretical and
  observational constraints on the HI intensity power spectrum}},
  \href{https://doi.org/10.1093/mnras/stu2702}{\emph{Mon. Not. Roy. Astron.
  Soc.} {\bfseries 447} (2015) 3745},
  [\href{https://arxiv.org/abs/1407.6366}{{\ttfamily 1407.6366}}].

\bibitem{Masui:2010mp}
K.~W. Masui, P.~McDonald and U.-L. Pen, \emph{{Near term measurements with 21
  cm intensity mapping: neutral hydrogen fraction and BAO at z<2}},
  \href{https://doi.org/10.1103/PhysRevD.81.103527}{\emph{Phys. Rev.}
  {\bfseries D81} (2010) 103527},
  [\href{https://arxiv.org/abs/1001.4811}{{\ttfamily 1001.4811}}].

\bibitem{Masui:2012zc}
K.~Masui, E.~Switzer, N.~Banavar, K.~Bandura, C.~Blake et~al.,
  \emph{{Measurement of 21 cm brightness fluctuations at z\textasciitilde0.8 in
  cross-correlation}},
  \href{https://doi.org/10.1088/2041-8205/763/1/L20}{\emph{Astrophys.J.}
  {\bfseries 763} (2013) L20},
  [\href{https://arxiv.org/abs/1208.0331}{{\ttfamily 1208.0331}}].

\bibitem{Wolz:2015lwa}
L.~Wolz et~al., \emph{{Erasing the Milky Way: new cleaning technique applied to
  GBT intensity mapping data}},
  \href{https://arxiv.org/abs/1510.05453}{{\ttfamily 1510.05453}}.

\bibitem{Parkinson:2012vd}
D.~Parkinson et~al., \emph{{The WiggleZ Dark Energy Survey: Final data release
  and cosmological results}},
  \href{https://doi.org/10.1103/PhysRevD.86.103518}{\emph{Phys. Rev.}
  {\bfseries D86} (2012) 103518},
  [\href{https://arxiv.org/abs/1210.2130}{{\ttfamily 1210.2130}}].

\bibitem{Amendola:2016saw}
L.~Amendola et~al., \emph{{Cosmology and Fundamental Physics with the Euclid
  Satellite}},  \href{https://arxiv.org/abs/1606.00180}{{\ttfamily
  1606.00180}}.

\end{thebibliography}\endgroup


\end{document}